\documentclass[prd,twocolumn,showpacs,preprintnumbers,amsmath,amssymb,floatfix,nofootinbib]{revtex4-1}

\usepackage{subfigure,graphicx,epsfig,amsmath,amsfonts,amssymb,xcolor,slashed,ulem}

\usepackage{bm}
\usepackage{graphicx}
\usepackage{amsmath}
\usepackage{amsfonts}
\usepackage{amssymb}
\usepackage{color}
\usepackage{ulem}
\usepackage{float}
\usepackage[stable]{footmisc}
\usepackage[section]{placeins}

\usepackage{booktabs}
\usepackage[colorlinks, citecolor=blue,anchorcolor=red,menucolor=red, linkcolor=red,filecolor=red,runcolor=red,urlcolor=blue,frenchlinks=red]{hyperref}
\usepackage{etoolbox}
\usepackage{orcidlink}  

\newcommand{\be}{\begin{equation}}
\newcommand{\ee}{\end{equation}}
\newcommand{\ba}{\begin{eqnarray}}
\newcommand{\ea}{\end{eqnarray}}
\newcommand{\nn}{\nonumber}
\newcommand{\ds}{D_{s1}}
\newcommand{\mev}{\textrm{ MeV}}

\begin{document}
\bibliographystyle{unsrt}
\arraycolsep1.5pt

\title{The $D_{s1}(2460) \to D_s \pi^+ \pi^- $ decay from a $D_{s1}$ molecular perspective}

\author{Luis Roca\orcidlink{0000-0001-8067-5320}}
\affiliation{\it  Departamento de F\'{\i}sica. Universidad de
Murcia. E-30100 Murcia, Spain}

\author{Jorgivan Morais Dias\orcidlink{0000-0002-0354-4711}}%
\affiliation{Departamento de F\'{\i}sica,  Universidade Federal do Piau\'{\i}, 64049-550 Teresina, Piau\'{\i}, Brasil}

\author{Eulogio Oset}
\affiliation{Department of Physics, Guangxi Normal University, Guilin 541004, China}
\affiliation{Departamento de F\'{\i}sica Te\'orica e IFIC, Centro Mixto
Universidad de Valencia-CSIC, Institutos de Investigacion de
Paterna, Apdo 22085, 46071 Valencia, Spain}

\begin{abstract}
We conduct a theoretical study of the $ D_{s1}(2460) \to D_s \pi^+ \pi^- $ 
decay from the perspective that the $D_{s1}$ is a molecular state, built 
mostly from the $D^* K$ and $D_s^* \eta$ components. The $D^*$ and $D_s^*$ 
mesons are allowed to decay into two pseudoscalars, with one 
of them merging with the other pseudoscalar that forms the $D_{s1}$ state, 
ultimately leading to the $\pi^+ \pi^- D_s$ final state. This results 
in a triangle diagram mechanism where all theoretical ingredients are 
well-known, leading to a free parameter framework. We evaluate the 
mass distributions of particle pairs and find good agreement with the 
experimental distributions of a recent LHCb experiment, providing 
strong support to the molecular picture of the $D_{s1}(2460)$ state.
We also discuss the role played by the scalar mesons $f_0(500)$ and 
$f_0(980)$, at odds with the interpretation of the experimental analysis.
\end{abstract}

\maketitle

\section{Introduction}
The $D_{s1}(2460)$ state is the subject of intense debate 
concerning its structure. It is well established that its mass 
is significantly lower than that predicted by standard quark 
models based on $q \bar{q}$ structure \cite{130,140,150}. The proximity of this state to the $D^*K$ threshold has 
prompted studies suggesting it could be predominantly a $D^*K$ molecular state, 
though coupled-channel dynamics will inevitably mix it with other channels 
\cite{8,9,10,11,gamermann,13,131,132,133,134,135,136,137,138,rosner,70,80}. 
The molecular nature is also supported within a chiral 
quark model calculation \cite{81}. Lattice QCD data also 
supports this picture \cite{sasa,bali}, and the scattering 
length and effective range obtained from the lattice data 
also suggest a largely molecular picture \cite{composi}. 
Tetraquark assignments have also been advocated \cite{14,15}. 
Some other works use $q \bar{q}$ components, with sophisticated 
potentials, or sum rules \cite{17,18,19,20}, and there are 
also works suggesting a mixture of these various configurations 
\cite{21,22,79,231,232}. A review paper on these issues is 
given in \cite{230}.

The present work investigates the $D_{s1}(2460) \to D_s \pi^+ \pi^- $ 
decay, which has been recently measured by the LHCb collaboration 
\cite{expe}. Theoretical work on this reaction, prior to its 
observation in \cite{expe}, is scarce. In \cite{45}, a calculation 
of the decay rate assuming a $c \bar{s}$ state was done, but 
assuming the $D_s$ at rest. In this scenario, the $\pi \pi$ system 
appears in p-wave to conserve angular momentum and thus the $\pi \pi$ 
system is in $I=1$, violating isospin in the decay. Yet, the reaction 
can proceed via pions in s-wave, as we shall see, and the decay mode 
does not need to violate isospin.
 A more detailed work, making predictions of mass distributions, 
 was done in \cite{hanhart}, assuming that the $D_{s1}$ state is a 
 molecular state of $D^* K$. A double hump structure was predicted 
 for the $\pi \pi$ distribution, which has indeed been observed in 
 the LHCb experiment \cite{expe}. Our work is similar to that of 
 \cite{hanhart}, and we also assume the $D_{s1}$ to be a molecular 
 state, yet we include also the $D_s^* \eta$ coupled channel, and the 
 work is technically quite different. Following the experiment, another 
 paper on the reaction appeared in \cite{wangliluo}, which assumes a 
 minimal $c \bar{s}$ structure for the $D_{s1}$ state and uses SU(3) 
 symmetry by looking at the corresponding matrix element of the 
 $\langle PPP \rangle$ structure, with $P$ the $q \bar{q}$ matrix 
 written in terms of mesons. That work, implementing the final state 
 interaction of the mesons reproduces the experimental mass 
 distributions at the cost of fitting five free parameters. Compared 
 to the work of \cite{hanhart}, that work misses the $D^* K$ structure 
 of the first decay of the $D_{s1}$, prior to any final state interaction.

Our work assumes the molecular structure of the $D_{s1}$ in coupled channels, 
and according to \cite{gamermann}, only the $D^* K$ and $D_s^* \eta$ channels 
are relevant. All the couplings and scattering amplitudes involved in the 
theoretical framework are well established, allowing us to evaluate the decay 
rate and mass distributions without the need for any free parameters.

\section{Formalism}

\subsection{The $\ds(2460)$ molecular picture
\label{sec:Ds1}}

Our model for the $\ds(2460)$  resonance relies on  
the picture of \cite{gamermann}, where it is dynamically generated from the coupled channels $DK^*$, $KD^*$, $\eta D_s^*$, $D_s \omega$, $\eta_c D_s^*$, $D_s J/\Psi$. The interaction between these channels is described by  transition potentials $V_{ij}$ obtained from effective Lagrangians and the on-shell scattering matrix is given by

\begin{eqnarray}
T = [1 - V G]^{-1} V \,,
\label{eqBS}
\end{eqnarray}
with $G$ a diagonal matrix containing the meson-meson intermediate loop functions, which can be regularized with a sharp cutoff or dimensional regularization. In the sharp cutoff method, $G$ is given by

\begin{eqnarray}
G=\int_{|\vec q|<q_{\text{max}}}\frac{d^3\vec q}{(2\pi)^3}\frac{\omega_1+\omega_2}{2\omega_1\omega_2}
\frac{1}{s-(\omega_1+\omega_2)^2+i\epsilon}\ ,
\label{eq:Gloop}
\end{eqnarray}
where $\omega_i=\sqrt{\vec q\,^2+m_i^2}$ and $q_{\text{max}}$ is the cutoff parameter, which, according to \cite{danijuan,jingdai}, stands for the range of the interaction in momentum space.
Indeed, as shown in Ref.~\cite{danijuan}, one can see that starting from a potential

\begin{eqnarray}
V(\vec p,\vec p\,')= \Theta(q_{\text{max}}-|\vec p|) 
\Theta(q_{\text{max}}-|\vec p\,'|)V,
\label{eq:Vcut}
\end{eqnarray}
one obtains Eqs.~\eqref{eqBS} and \eqref{eq:Gloop}, and it also leads to the momentum dependence for the scattering amplitude $T$
\begin{eqnarray}
T(\vec p,\vec p\,')= \Theta(q_{\text{max}}-|\vec p|)\Theta(q_{\text{max}}-|\vec p\,'|)T,
\label{eq:Tcut}
\end{eqnarray}
with $T$ given in Eq.~\eqref{eqBS}.
Equation \eqref{eq:Tcut} plays a crucial role in our approach, as it naturally regularizes the loop integrals that arise in our formalism.

A natural size of $q_{\text{max}}$ is of the order of the exchanged vector meson, as this determines the scale of the range of the potential. In Ref.~\cite{135}, values around $q_{\text{max}} \sim 800$~MeV  were used in order to reproduce the pole of the $\ds$ at the right position. Consequently, in the present work, we take $q_{\text{max}} \simeq 800$~MeV, which will later be used in the evaluation of the triangle loops, as explained in section~\ref{sec:loopeval}. We have checked that variations in the cutoff within the range allowed in Ref.~\cite{135} lead to uncertainties of only a few percent in the final results.

In \cite{gamermann}, it was found that the coupled channels mentioned above generate two states, one around 2456~MeV, which was associated to the $D_{s1}(2460)$, and another one at 2574~MeV, associated with the $D_{s1}(2536)$, both with quantum numbers $J^P = 1^+$. The $D_{s1}$ resonance was found to couple strongly to $KD^*$, with a coupling strength of $|g_{K D_s^*}| \simeq 10$ GeV, and to $\eta D_s^*$ with a coupling of $|g_{KD_s^*}| \simeq 6$ GeV.
Using our isospin phase convention for the multiplets $(D^+,-D^0)$, $(D^{*+},-D^{*0})$ and $(K^+,K^0)$, the relative sign between these two couplings is negative, as found in \cite{ikeno}. In Ref.~\cite{jiaxin}, using an extension of the local hidden gauge approach \cite{hidden1,hidden2,hidden4,hideko}, where the interaction stems from the exchange of vector mesons, the coupling $|g_{KD^*}|=12.11$~GeV was obtained. 
In the present work we take the following values for  the couplings:
 $g_{KD^*} = 12.11$ GeV, $g_{\eta D_s^*} = -6.16$ GeV,
and considering  that the $I = 0$ state of the $D_{s1}$ is $|D^* K, I = 0\rangle = \frac{1}{\sqrt{2}}(D^{*+}K^0 + D^{*0}K^+)$, we have $g_{D^{*0} K^+} = \frac{1}{\sqrt{2}} g_{D^* K}$, $g_{D^{*+} K^0} = \frac{1}{\sqrt{2}} g_{D^* K}$. The vertex $D_{s1} \to D^* K$ is then given by
\begin{eqnarray}
V_{D_{s1}, D^* K}(\vec{p}\,') = g_{D^* K} \, \vec{\epsilon}_A \cdot \vec{\epsilon}_{D^*} \, \Theta(q_{\text{max}} - |\vec{p}\,'|),
\end{eqnarray} which gives rise to the following  $D^* K \to D^* K$ amplitude, corresponding to the diagram shown in Fig.~\ref{fig:diagDKtree}:

\begin{figure}[h]
\begin{center}
\includegraphics[width=0.25\textwidth]{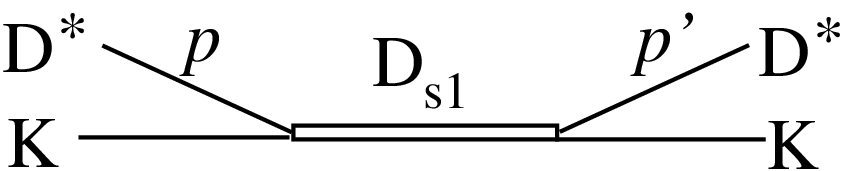}
\caption{\small{Amplitude for $D^*K \to D^* K$ mediated by the $\ds$ resonance.}}
\label{fig:diagDKtree}
\end{center}
\end{figure}

\begin{align}
&t_{D^* K, D^* K}(M_{\text{inv}}, \vec{p}, \vec{p}\,') = \frac{g_{D^* K}^2}{M_{\text{inv}}^2 - M_{D_{s1}}^2} \nonumber \\
&\times \sum_i \epsilon_{A_i} \epsilon'_{D^* i} \epsilon_{A_j} \epsilon_{D^* j} \, \Theta(q_{\text{max}} - |\vec{p}|) \, \Theta(q_{\text{max}} - |\vec{p}\,'|) \nonumber \\
&= \frac{g_{D^* K}^2}{M_{\text{inv}}^2 - M_{D_{s1}}^2}
\vec{\epsilon}\,'_{D^*} \cdot \vec{\epsilon}_{D^*} \, \Theta(q_{\text{max}} - |\vec{p}\,|)\Theta(q_{\text{max}} - |\vec{p}\,'|).
\label{eq:tDstarK}
\end{align}
This gives the structure of the s-wave vector-pseudoscalar amplitudes in \cite{gamermann}.
It is worth stressing that Fig.~\ref{fig:diagDKtree}, and its corresponding parametrization in Eq.~\eqref{eq:tDstarK}, represents an effective description of the amplitude in Eq.~\eqref{eqBS} near its poles. Indeed, in Eq.~\eqref{eqBS}, the $\ds$ emerges dynamically by the highly non-linear dynamics involved in that equation without the need to include it as an explicit degree of freedom.

\subsection{The $D_{s1}(2460) \to D_s \pi^+ \pi^- $ formalism}

Once the model for the dynamical generation of the $\ds$ has been established, 
we now proceed to explain the mechanisms responsible for the $D_{s1}(2460) \to D_s \pi^+ \pi^- $ decay amplitude in our approach, which are depicted in Fig.~\ref{fig:diagtriang}.

\subsubsection{$VPP$ vertices}

\begin{figure*}[htbp]
\begin{center}
\includegraphics[width=0.85\textwidth]{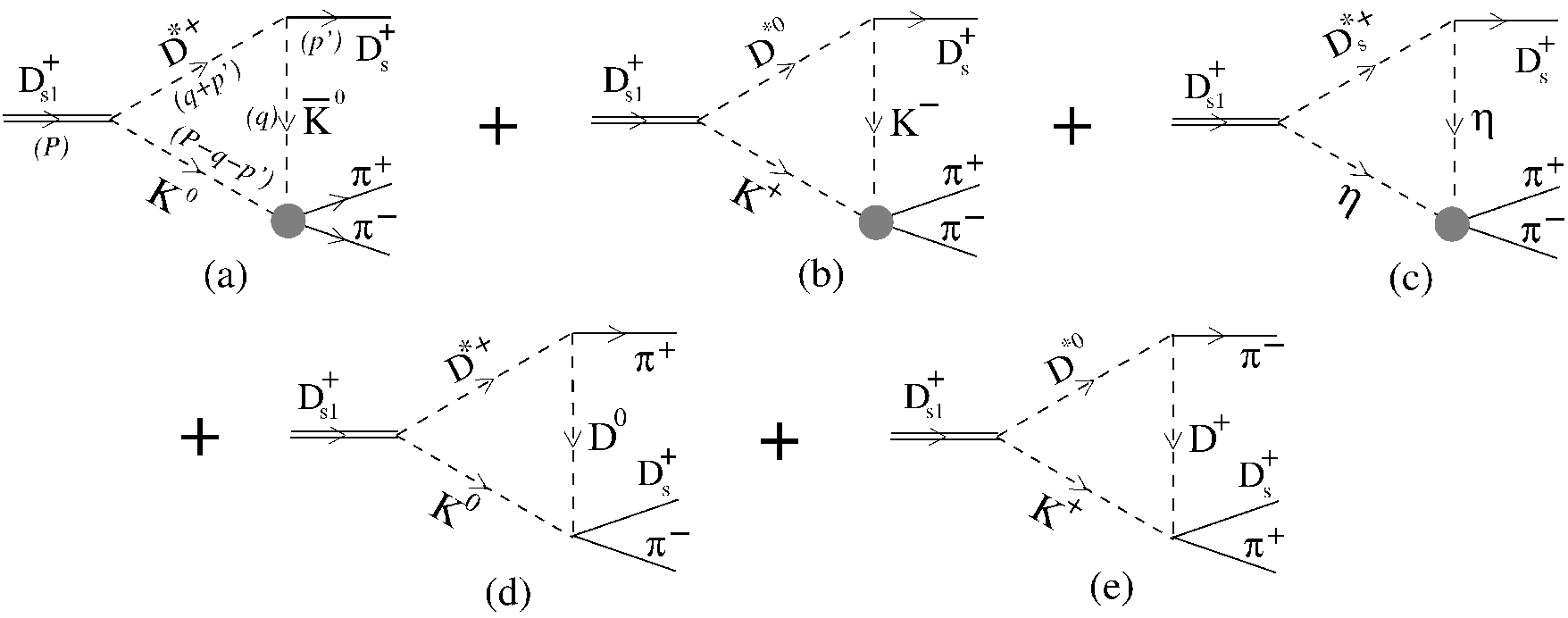}
\caption{\small{Feynman diagrams involved in the $D_{s1} \to D_s \, \pi^+ \pi^-$ decay from the $D_{s1}$ molecular picture, with the momenta of the particles indicated in parentheses.}}
\label{fig:diagtriang}
\end{center}
\end{figure*}

First, we need the vector ($V$) to two pseudoscalar ($P$) vertices, which are given by the Lagrangian \cite{hideko}

\begin{eqnarray}
\mathcal{L}_{VPP} = -i g\langle [P, \partial_\mu P] V^\mu \rangle,
\label{eq:LVPP}
\end{eqnarray}
with $g = \frac{M_V}{2f}$, $M_V = 800$~MeV and $f = 93$~MeV, while $P$, $V^\mu$ are the matrices of $q\bar{q}$ written in terms of mesons, given by

\begin{eqnarray}
 P = \left( \begin{array}{cccc}
 \frac{\pi^0}{\sqrt{2}} + \frac{\eta}{\sqrt{3}} & \pi^+ & K^+ & \bar{D}^0 \\
 \pi^- & -\frac{\pi^0}{\sqrt{2}} + \frac{\eta}{\sqrt{3}} & K^0 & D^- \\
 K^- & \bar{K}^0 & -\frac{\eta}{\sqrt{3}} & D_s^- \\
 D^0 & D^+ & D_s^+ & \eta_c \\
 \end{array} \right) 
 \label{eqPmatrix}
\end{eqnarray}

\begin{eqnarray}
 V = \left( \begin{array}{cccc}
 \frac{\rho^0}{\sqrt{2}} + \frac{\omega}{\sqrt{2}} & \rho^+ & K^{*+} & \bar{D}^{*0} \\
 \rho^{*-} & -\frac{\rho^0}{\sqrt{2}} + \frac{\omega}{\sqrt{2}} & K^{*0} & D^{*-} \\
 K^{*-} & \bar{K}^{*0} & \phi & D_{s}^{*-} \\
 D^{*0} & D^{*+} & D_{s}^{*+} & J/\psi \\
 \end{array} \right)\, ,
 \label{vecfie}
\end{eqnarray}
where we have assumed the standard $\eta-\eta'$ physical
mixing between the singlet and octet SU(3) states \cite{bramon}.

From this Lagrangian, and considering the momenta of the particles shown in Fig.~\ref{fig:diagtriang}, the $VPP$ vertices take the form:

\begin{equation}
t_{VPP,j} = \beta_j\,g \,h\, \vec{\epsilon}\cdot (\vec{q} - \vec{p}\,') \, ,
\label{eqtVPP}
\end{equation}
with $\beta_j$ given in Table~\ref{tab:beta} for the different vertices,
\begin{table}[ht]
\centering
\resizebox{\columnwidth}{!}{%
\begin{tabular}{|c|c|c|c|c|c|}
\hline
 & \(D^{*+} \to D_s^+ \bar{K}^0\) & \(D^{*0} \to D_s^+ K^-\) & \(D_s^{*+} \to D_s^+ \eta\) & \(D^{*+} \to D^0 \pi^+\) & \(D^{*0} \to D^+ \pi^-\) \\
\hline
$\beta_j$ & \(-1\) & \(-1\) & \(\frac{1}{\sqrt{3}}\) & \(1\) & \(1\) \\
\hline
\end{tabular}
}
\caption{Values of \( \beta_j \) for the different $VPP$ vertices.}
\label{tab:beta}
\end{table}
and $h\equiv \frac{m_{D^*}}{m_{K^*}}$ is a factor required by heavy quark symmetry, as explained in a similar context in Ref.~\cite{Liang:2014eba}.

For the $D^{*}D\pi$ vertices, which appear in the diagrams in Fig.~\ref{fig:diagtriang}(d) and (e), we can take advantage of the available experimental data for the decay width  $\Gamma(D^{*0} \to D^+\pi^-)=56.5$~keV \cite{pdg}. Evaluating this decay with the amplitude of Eq.~\eqref{eqtVPP}, we get
\begin{align}
\Gamma(D^{*0} \to D^+\pi^-)=\frac{g^2\,h^2}{6\pi\, m^2_{D^{*}}} q^3=75.1~\textrm{keV}\, .
\end{align}
To ensure consistency with the experimental value, we fine-tune the $D^{*}D\pi$ vertices in our model by multiplying them by a factor of $0.867$, as would be requested to get the experimental 
$D^{*0} \to D^+\pi^-$ decay width. In the following equations, this factor is implicitly assumed in all amplitudes involving this vertex.

\subsubsection{The $K \bar{K} \to \pi \pi$, $\eta \eta \to \pi \pi$ amplitudes
\label{secTunit}}

In the lower vertices of the diagrams in Fig.~\ref{fig:diagtriang}(a), (b), and (c), represented by the thick dots, the pseudoscalar mesons inside the triangle loop undergo final state interaction (FSI) in coupled channels, that leads to the production of the final $\pi^+\pi^-$ pair.
To evaluate these meson-meson scattering amplitudes,
we use the chiral unitary approach of \cite{npa}, which accounts for the loop resummation in s-wave in an analogous way to Eq.~\eqref{eqBS}.
 Specifically, we include the $K^0 \bar{K}^0$, $K^+ K^-$, $\pi^+ \pi^-$, $\pi^0 \pi^0$, $\eta \eta$ as coupled channels with the transition potentials given in \cite{liangoset}. Alternatively,  the slightly modified potentials of \cite{linliang} can be used, where the 
 $\eta$-$\eta'$ mixing of \cite{bramon} is also considered, but the differences are minimal.
Since the $\pi\pi$ state is external, we require the on-shell amplitude in s-wave as a function of $M_{\text{inv}}(\pi^+ \pi^-)$, and thus it goes out of the loop integral. We refer the reader to these references for further details and explanations.
These unitarized meson-meson scattering amplitudes generate dynamically the scalar resonances, which appear as poles in unphysical Riemann sheets. Importantly, the unitarized scattering amplitudes account not only for the resonant behavior but for the full scattering amplitude. The model relies on a single free parameter, the regularization of the meson-meson loop function, for which we use a three-momentum cutoff value of 600~MeV, as in \cite{linliang}, which reproduces well the scattering data at the energies relevant in the present work.

\subsubsection{The $K^0 D^0 \to \pi^- D_s^+$, and $K^+ D^+ \to \pi^+ D_s^+$ amplitudes}

\begin{figure}[h]
\begin{center}
\includegraphics[width=0.45\textwidth]{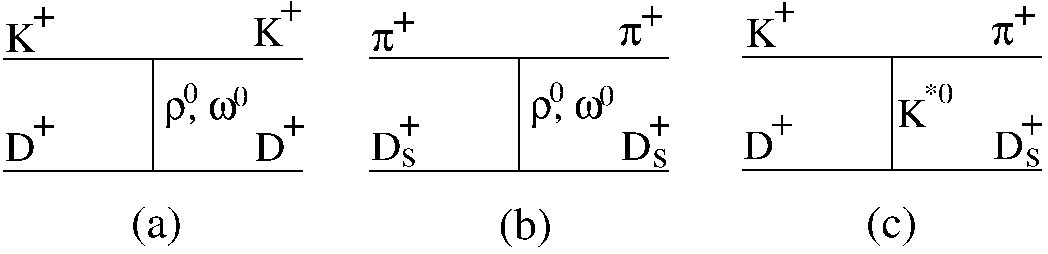}
\caption{\small{Vector meson exchange potentials needed in the evaluation of the $D K \to D_s \pi$ amplitudes .}}
\label{diagsDKtreetres}
\end{center}
\end{figure}

For the lower vertices in the diagrams of Fig.~\ref{fig:diagtriang}(d) and (e), we need the $K^0 D^0 \to \pi^- D_s^+$ and $K^+ D^+ \to \pi^+ D_s^+$ amplitudes, which have not been previously evaluated within the chiral unitary approach. These amplitudes can be obtained using the extended local hidden gauge approach, exchanging vector mesons. We have to consider the coupled channels $K^+ D^+$ and  $\pi^+ D_s^+$, which are $I=1$, $I_3=+1$, states. The corresponding potentials are represented diagrammatically in Fig.~\ref{diagsDKtreetres}.

The evaluation of these potentials is straightforward from the Lagrangian of Eq.~\eqref{eq:LVPP} for both vertices, and we find:

\begin{align}
&V_{K^+ D^+, K^+ D^+} = 0; \quad
V_{\pi^+ D_s^+, \pi^+ D_s^+} = 0; \nn \\
&V_{K^+ D^+, \pi^+ D_s^+} = \frac{g^2}{m^2_{K^*}} ({p_1} + {p_3})_\mu  ({p_2} + {p_4})^\mu.
\end{align}

The null result for diagram \ref{diagsDKtreetres}(a) stems from a cancellation between $\rho^0$ and $\omega$ exchange. The zero for diagram \ref{diagsDKtreetres}(b) is readily explained, as there is no overlap between the $u$, $d$, quarks of the pions and the $c$, $\bar{s}$, of the $D_s$. The non-diagonal potential can be projected over $s$-wave and we find \cite{luissingh}

\begin{align}
V_{K^+ D^+, \pi^+ D_s^+}(s) &= \frac{g^2}{2m^2_{K^*}} \left[ 3s - \left( m_{K^+}^2 + m_{\pi^+}^2 + m_D^2 + m_{D_s}^2 \right) \right. \nonumber \\
& \quad - \frac{1}{s} \left( m_D^2 - m_{K^+}^2 \right) \left( m_{D_s}^2 - m_{\pi^+}^2) \right].
\label{eq:VKD}
\end{align}

Due to isospin symmetry, the $V_{K^0 D^0, \pi^- D_s^+}$ transition ($I=1$, $I_3=-1$) is  the same as $V_{K^+ D^+, \pi^+ D_s^+}$.  (Note particularly that the phases from the isospin multiplet ($-\pi^+, \pi^0, \pi^-$) and the $D^0$ make the sign equal in both cases). Since the diagonal transitions in Fig.~\ref{diagsDKtreetres} are zero, there is no need to work with coupled channels. Furthermore the  $K^0 D^0 \to \pi^- D_s^+$ and $K^+ D^+ \to \pi^+ D_s^+$ amplitudes are non-resonant, and thus it suffices to consider only the lowest order tree level potentials, and thus the $K^+ D^+ \to \pi^+ D_s^+$ transition amplitude is given in Eq.~\eqref{eq:VKD}, and the same for  $K^0 D^0 \to \pi^- D_s^+$.

\subsubsection{Loop evaluation
\label{sec:loopeval}
}

We are now ready to evaluate the triangle loops in Fig.~\ref{fig:diagtriang}. We explicitly explain the evaluation of the amplitude for the diagram of Fig.~\ref{fig:diagtriang}(a), from which the others can be readily deduced. Using the amplitudes for the vertices obtained in the previous sections, we find

\begin{align}
& -it^{(a)} = -i g_{D^{*+} K^0} \, \vec{\epsilon}_A \cdot \vec{\epsilon} \, \int \frac{d^4q}{(2\pi)^4} \frac{i}{q^2 - m_K^2 + i\varepsilon} \nn \\
& \times \frac{i}{(q + p')^2 - m_{D^*}^2 + i\varepsilon} \frac{i}{(P - q - p')^2 - m_K^2 + i\varepsilon} \nn \\
& \times i \, \vec{\epsilon} \cdot (\vec{q} - \vec{p}\,') \, g\,h \, \, (-i) \Theta(q_{\text{max}}-|\vec q+\vec p\,'|)t_{\bar{K}^0 K^0, \pi^+ \pi^-},
\end{align}
where $t_{\bar{K}^0 K^0, \pi^+ \pi^-}$ is the unitarized $\bar{K}^0 K^0\to \pi^+ \pi^-$ scattering amplitude described in section~\ref{secTunit}.

Summing over the $ D^* $ polarization,
$
\sum_{\text{pol}} \epsilon_i \epsilon_j = \delta_{ij},
$
we get

\begin{align}
&t^{(a)} =  -ig\,h \, g_{D^{*+} K^0} \, t_{\bar{K}^0 K^0, \pi^+ \pi^-} \epsilon_{Aj} \nn \\
& \times \int \frac{d^4q}{(2\pi)^4} \frac{1}{q^2 - m_K^2 + i\varepsilon}
 \frac{1}{(q + p')^2 - m_{D^*}^2 + i\varepsilon} \nn \\
 & \times \frac{1}{(P - q - p')^2 - m_K^2 + i\varepsilon} (q - p')_j \, \Theta(q_{\text{max}}-|\vec q+\vec p\,'|).
 \label{eq:ta2}
\end{align}

Next, we take advantage  of separating the positive energy part from the negative energy one in a meson propagator, which then reads 
\begin{align}
\frac{1}{k^2 - m_i^2 + i\varepsilon} &= \frac{1}{2 \omega_i(k)} \left( \frac{1}{k^0 - \omega_i(k) + i \varepsilon} \right. \nn \\
&\quad \left. - \frac{1}{k^0 + \omega_i(k) - i \varepsilon} \right)\, ,
\end{align}
with $\omega_i(k)=\sqrt{\vec k\,^2+m_i^2}$. Since in the decay of $D_{s1}$ to $D^* K^0$ both particles are close to being on-shell, it is sufficient to keep the positive energy part for these two particles, but we keep the two terms for the $\bar{K}^0$ propagator since it can be far off-shell. By focusing only on the integral in the expression of $t^{(a)}$ in Eq.~\eqref{eq:ta2}, we find

\begin{eqnarray}
Y_j &=& i \int \frac{d^4q}{(2\pi)^4} \frac{1}{2\omega_K(q)} \frac{1}{2\omega_{D^*}(\vec{q} + \vec{p}\,')} \frac{1}{2\omega_K(\vec{q} + \vec{p}\,')} \nonumber \\
&& \times \frac{1}{q^0 + p'^0 - \omega_D^*(\vec{q} + \vec{p}\,') + i \varepsilon} \nn\\
&& \times \frac{1}{P^0 - q^0 - p'^0 - \omega_K(\vec{q} + \vec{p}\,') + i \varepsilon} \nonumber \\
&& \times \left[ \frac{1}{q^0 - \omega_K(q) + i \varepsilon} - \frac{1}{q^0 + \omega_K(q) - i \varepsilon} \right] \nonumber \\
&& \times \Theta(q_{\text{max}} - |\vec{q} + \vec{p}\,'|) \, (p' - q)_j.
\end{eqnarray}

The $q^0$ integration is immediately done using Cauchy integration, evaluating residues, and we find

\begin{eqnarray}
Y_j &=& \int \frac{d^3q}{(2\pi)^3} \frac{1}{2\omega_K(q)} \frac{1}{2\omega_{D^*}(\vec{q} + \vec{p}\,')} \frac{1}{2\omega_K(\vec{q} + \vec{p}\,')} \nonumber \\
&& \times \frac{1}{P^0 - \omega_{D^*}(\vec{q} + \vec{p}\,') -  \omega_K(\vec{q} + \vec{p}\,') + i \varepsilon} \nn \\
&& \times \left[ \frac{1}{P^0 - p'^0 - \omega_K(\vec{q} + \vec{p}\,') - \omega_K(q) + i \varepsilon} \right. \nonumber \\
&& \left. + \frac{1}{p'^0 - \omega_{D^*}(\vec{q} + \vec{p}\,') - \omega_K(q) + i \varepsilon} \right] \nonumber \\
&& \times \Theta(q_{\text{max}} - |\vec{q} + \vec{p}\,'|) (p'-q)_j,
\end{eqnarray}
and taking into account that 

\begin{equation}
\int d^3q \, q_j \, f(\vec{q}, \vec{p}\,') = p'_j \int d^3q \, \frac{\vec{q} \cdot \vec{p}\,'}{|\vec{p}\,'|^2} \, f(\vec{q}, \vec{p}\,'),
\end{equation}
we finally find
\begin{align}
t^{(a)} =  g\,h \, g_{D^{*+} K^0} \, t_{\bar{K}^0 K^0, \pi^+ \pi^-} \vec \epsilon_{A} \cdot\vec p\,'\, Y(D^*,K,\bar K)\, ,
\end{align}
with

\begin{eqnarray} 
&&Y(D^*,K,\bar{K}) = \int \frac{d^3q}{(2\pi)^3} 
\left(1 - \frac{\vec{q} \cdot \vec{p}\,'}{|\vec{p}\,'|^2} \right) \nn\\
&& \times
\frac{1}{2\omega_K(q)} \frac{1}{2\omega_{D^*}(\vec{q} + \vec{p}\,')} \frac{1}{2\omega_K(\vec{q} + \vec{p}\,')} \nonumber \\
&& \times \frac{1}{P^0 - \omega_{D^*}(\vec{q} + \vec{p}\,') -  \omega_K(\vec{q} + \vec{p}\,') + i \varepsilon} \nonumber \\
&& \times \left[ \frac{1}{P^0 - p'^0 - \omega_K(\vec{q} + \vec{p}\,') - \omega_K(q) + i \varepsilon} \right. \nonumber \\
&& \left. + \frac{1}{p'^0 - \omega_{D^*}(\vec{q} + \vec{p}\,') - \omega_K(q) + i \varepsilon} \right]\nn\\
&& \times \Theta(q_{\text{max}} - |\vec{q} + \vec{p}\,'|)\, . 
\label{eq:YNTK}
\end{eqnarray}

By an argument analogous to the one leading to the appearance of the $\Theta$ step function in the integral of Eq.~\eqref{eq:YNTK}, which comes from the $\Theta$ function in Eq.~\eqref{eq:tDstarK}, we must also introduce an additional three-momentum cutoff associated with the amplitudes at the lower vertices in  Fig.~\ref{fig:diagtriang}. 
Indeed, similarly to the structure of Eq.~\eqref{eq:tDstarK} with two $\Theta$ functions, the chiral unitary approach with the cutoff regularization can be deduced by a separate potential, $V(\vec q,\vec q\,')=V \Theta(q_{\text{max}}-|\vec q|) \Theta(q_{\text{max}}-|\vec q\,'|)$, which reverts into $T(\vec q,\vec q\,')=T \Theta(q_{\text{max}}-|\vec q|) \Theta(q_{\text{max}}-|\vec q\,'|)$, where $q_{\text{max}}$ is the same one used to regularize the $G$ function in Eq.~\eqref{eq:Gloop} \cite{danijuan}.
Thus we must limit the momentum entering the lower vertices in Fig.~\ref{fig:diagtriang} with the same cutoff, which is $\Lambda\equiv 600\mev$ as explained in section~\ref{secTunit}.  For diagrams \ref{fig:diagtriang} (d) and (e), since the $DK\to D_s \pi$ vertex could,  in principle, have a different cutoff (denoted as  $\Lambda'$), we allow it to vary within the range $\Lambda'\in[500,700]$ to estimate the uncertainty in our calculation.  Since $\Lambda$ (or $\Lambda'$) corresponds to the three-momentum in the final $PP$ rest frame, we must boost the  momentum $q$ into this frame, yielding
$
{q^*}=({{q^*}^0}^2-m_{i}^2)^{1/2},
$
with
$
{q^*}^0=[(P^0-p'^0)\omega_i+ \vec q\cdot \vec p\,']/M_{PP},
$
with $m_i$ the mass of the pseudoscalar with momentum $q$  in the triangle loop, and $M_{PP}$ the invariant mass of the final $PP$ pair of the vertex. Consequently, the $\Theta$ function to be included in the integrand of Eq.~\eqref{eq:YNTK} is $\Theta(\Lambda-q^*)$, in addition to the one from the $D_{s1}\to D^* K$ already considered. 


The amplitude corresponding to the first three diagrams of Fig.~\ref{fig:diagtriang}(a), (b) and (c), can then be written as

\begin{equation}
t_K = \tilde{t}_K \, \vec{\epsilon}_A \cdot \vec{p}\,',
\end{equation}

with $\tilde{t}_K$ given by

\begin{eqnarray}
\label{eqtabc}
&&\tilde{t}_K =  \\
&& g \,h \, g_{D^{*+}K^0} \, t_{K^0\bar{K}^0,\pi^+\pi^-}\left(M_{\text{inv}}(\pi^+\pi^-) \right) \, Y(D^*,\bar{K},K) \nonumber \\
&& + g \,h \, g_{D^{*0}K^+} \, t_{K^+K^-,\pi^+\pi^-}\left(M_{\text{inv}}(\pi^+\pi^-)\right) \, Y(D^*,\bar{K},K) \nonumber \\
&& - \frac{1}{\sqrt{3}} g \,h \, g_{D_s^+\eta} \, t_{\eta\eta,\pi^+\pi^-}\left(M_{\text{inv}}(\pi^+\pi^-)\right) \, Y(D_s^+,\eta,\eta). \nn
\end{eqnarray}

The diagrams \ref{fig:diagtriang}(d) and (e) are calculated analogously, but now one is proportional to 
$\vec{\epsilon}_A \cdot \vec{p}_{\pi^+}$ and the other to 
$\vec{\epsilon}_A \cdot \vec{p}_{\pi^-}$. 
We find for diagrams 2(d),  2(e), respectively:

\begin{align}
t_D(\pi^+) &= \tilde{t}_D(\pi^+) \, \vec{\epsilon}_A \cdot \vec{p}_{\pi^+}, \\
t_D(\pi^-) &= \tilde{t}_D(\pi^-) \, \vec{\epsilon}_A \cdot \vec{p}_{\pi^-}.
\end{align}
with
\begin{align}
\tilde{t}_D(\pi^+) =&  
-\frac{g\,h}{\sqrt{2}} \, g_{D^{*}K} \, V_{K^+D^+, \pi^+D_s^+} \nn\\
& \times Y'(D^*,K,D;\pi^+),
\end{align}
and
\begin{eqnarray}
&&Y'(D^*,K,D;\pi^+) = \int \frac{d^3q}{(2\pi)^3} 
\left(1 - \frac{\vec{q} \cdot \vec{p}_{\pi^+}}{|\vec{p}_{\pi^+}|^2} \right) \nonumber \\ 
&& \times \frac{1}{2\omega_D(q)} \frac{1}{2\omega_{D^*}(\vec{q} + \vec{p}_{\pi^+})}  
\frac{1}{2\omega_K(\vec{q} + \vec{p}_{\pi^+})} \nonumber \\ 
&& \times \frac{1}{P^0 - \omega_{D^*}(\vec{q} + \vec{p}_{\pi^+})  
-  \omega_K(\vec{q} + \vec{p}_{\pi^+}) + i \varepsilon} \nonumber \\ 
&& \times \left[ \frac{1}{P^0 - p_{\pi^+}^0 - \omega_K(\vec{q} + \vec{p}_{\pi^+})  
- \omega_D(q) + i \varepsilon} \right. \nonumber \\ 
&& \left. + \frac{1}{p_{\pi^+}^0 - \omega_{D^*}(\vec{q} + \vec{p}_{\pi^+})  
- \omega_D(q) + i \varepsilon} \right] \nonumber \\ 
&& \times \Theta(q_{\text{max}} - |\vec{q} + \vec{p}_{\pi^+}|)
\Theta(\Lambda'-q^*).
\end{eqnarray}

The expression for 
$\tilde{t}_D(\pi^-)$ is the same as that for $\tilde{t}_D(\pi^+)$ simply exchanging $\pi^+$ and $\pi^-$.

Altogether we have, ($\vec{p}_{Ds}\equiv\vec{p}\,'  $),

\begin{equation}
    t^{\text{tot}} = \tilde{t}_K \vec{\epsilon}_A \cdot \vec{p}_{Ds} + \tilde{t}_D(\pi^+) \vec{\epsilon}_A \cdot \vec{p}_{\pi^+} + \tilde{t}_D(\pi^-) \vec{\epsilon}_A \cdot \vec{p}_{\pi^-}.
\end{equation}

Note that $|\vec{p}_{D_s}|$, $|\vec{p}_{\pi^+}|$ and $|\vec{p}_{\pi^-}|$ can be calculated in terms of $M_{\text{inv}}(ij)$ ($i,j \equiv 1,2,3$, in the order $D_s$(1), $\pi^-$(2), $\pi^+$(3)), but only two of these invariant masses are independent, since

\begin{equation}
M_{12}^2 + M_{13}^2 + M_{23}^2 = M_{D_{s1}}^2 + m_{D_s}^2 + m_\pi^2 + m_{\pi}^2.
\end{equation}

The double differential mass distribution is obtained as

\begin{equation}
\frac{d\Gamma}{dm_{12} dm_{23}} = \frac{1}{(2\pi)^3} \frac{2 m_{12} 2m_{23}}{32 m_{D_{s1}}^3} \overline{\sum} |t^{\text{tot}}|^2,
\label{eqdGmm}
\end{equation}
where the average over polarizations of the $D_{s1}$ is done in  $\overline{\sum}|t_{\text{tot}}|^2$, given by

\begin{align}
\overline{\sum} \sum |t^{\text{tot}}|^2 &= \frac{1}{3} \left\{ |\tilde{t}_K|^2 \vec{p}\,^2_{D_s^+} + |\tilde{t}_D(\pi^+)|^2 \vec{p}\,^2_{\pi^+} \right. \nn \\
&  + |\tilde{t}_D(\pi^-)|^2 \vec{p}\,^2_{\pi^-} + 2 \, \text{Re} \left( \tilde{t}_K \tilde{t}_D^*(\pi^+) \right) \vec{p}_{D_s^+} \cdot \vec{p}_{\pi^+} \nn \\
&  + 2 \, \text{Re} \left( \tilde{t}_K \tilde{t}_D^*(\pi^-) \right) \vec{p}_{D_s^+} \cdot \vec{p}_{\pi^-} \nn \\
& \left. + 2 \, \text{Re} \left( \tilde{t}_D(\pi^+) \tilde{t}_D^*(\pi^-) \right) \vec{p}_{\pi^+} \cdot \vec{p}_{\pi^-} \right\}\,.
\end{align}

Note that all the momenta and scalar products of momenta can be expressed in terms of the invariant masses, and hence $\overline{\sum} |t^{\text{tot}}|^2$ only depends on two independent invariant masses.

 The two-particle mass distributions can be obtained by integrating 
 Eq.~\eqref{eqdGmm} over one of the invariant masses within the limits given in the PDG \cite{pdg}. 
By cyclical permutation of these expressions, we can obtain all the mass distributions for the various final two-particle cases.

\section{Results}

\begin{figure}[htbp]
\begin{center}
\includegraphics[width=0.45\textwidth]{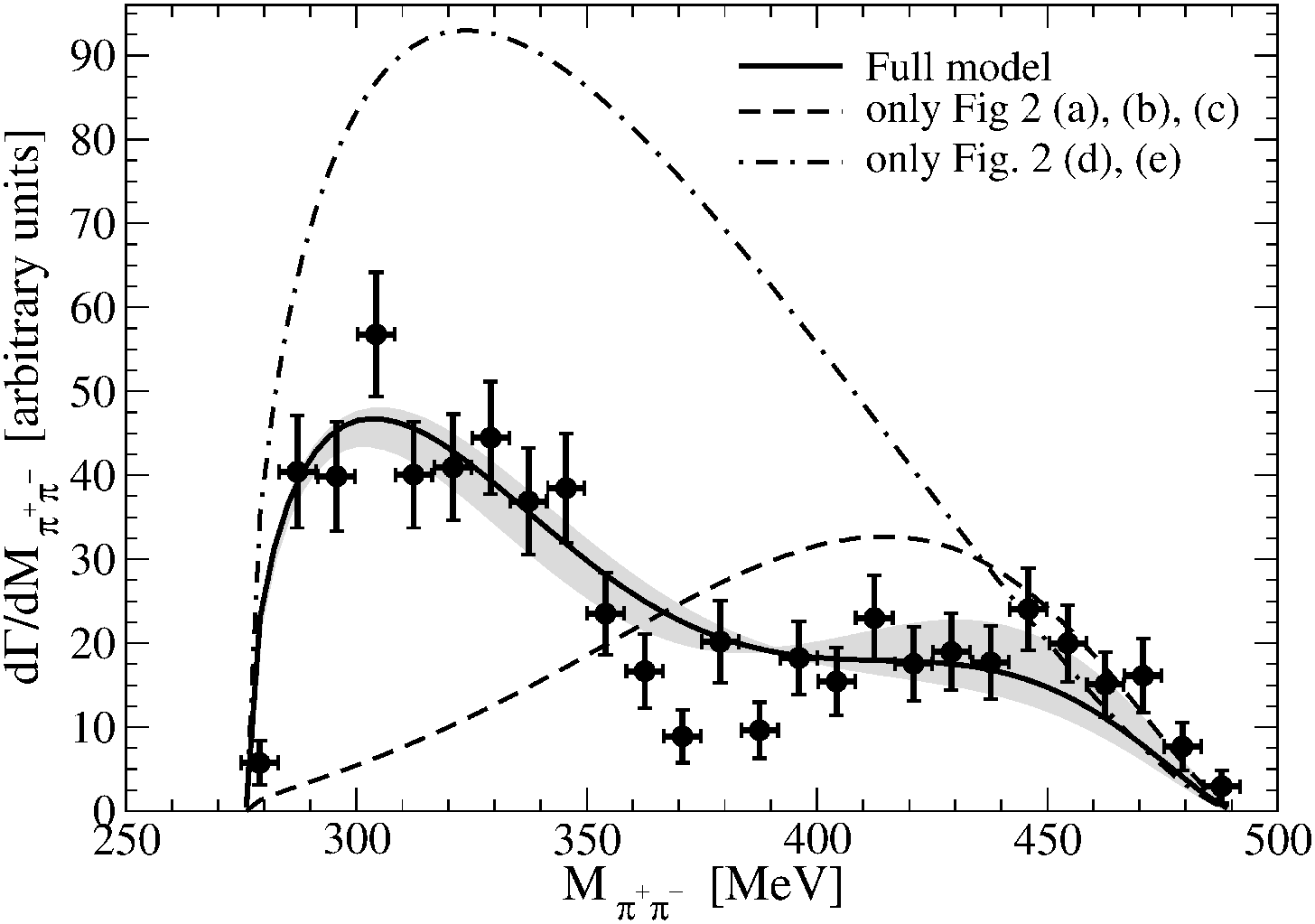}
\caption{\small{Comparison of the different contributions to the experimental $\pi^+\pi^-$ distribution \cite{expe}. The error band reflects the uncertainty from variations in the $\Lambda'$ cutoff parameter in the range $500-700$~MeV.}}
\label{fig:Minv_pipi}
\end{center}
\end{figure}

In Fig.~\ref{fig:Minv_pipi}, we show the results for the  $\pi^+\pi^-$ invariant mass distribution compared with the experimental results from \cite{expe}.  Since the experimental analysis does not provide an absolute normalization, we have scaled our full model result to match the area of the $\pi^+\pi^-$ distribution. However, our model does indeed predict the absolute strength, as will be discussed later. 
The error band reflects the uncertainty due to variations in the $\Lambda'$ cutoff parameter within the range $500-700$~MeV, while maintaining the normalization to the experimental area. This variation accounts for most of the uncertainty in the shape of the distribution within our model.
The agreement with the experiment is notably good, given the experimental uncertainties. Remarkably, this agreement arises from a nontrivial interference between diagrams  (a), (b), (c) with diagrams (d) and (e) of Fig.~\ref{fig:diagtriang}. 
We see that the contribution of the latter two diagrams is bigger than that of the first three, despite involving a $D$ propagator rather than the lighter $K$ or $\eta$. The reason is that the $D^* \to D\pi$ decay is kinematically allowed, bringing the intermediate $D$ in the loop close to on-shell, while the $D^* \to D_s^+ K$ is kinematically forbidden from going on-shell.

On the other hand, for the first three diagrams, which involve the transitions $K\bar{K} \to \pi^+\pi^-$ and $\eta\eta \to \pi^+\pi^-$, the amplitudes generate dynamically the $f_0(500)$ and $f_0(980)$ contributions, as mentioned above. In principle, the $f_0(500)$ should be suppressed relative to the $f_0(980)$ because the latter predominantly couples to $K\bar{K}$, while the coupling of $f_0(500)$ to $K\bar{K}$ is small \cite{npa}. However, in this reaction, the $f_0(980)$ is very far away from the region of $\pi^+\pi^-$ invariant masses allowed by the phase space, such that its contribution is practically absent in the figure. 
This contrasts significantly with one of the interpretations of the experimental analysis of \cite{expe}, where in one of the options the $f_0(980)$ contribution is found to be huge and responsible, through interference with the $f_0(500)$, of the experimental shape. 
 We also note here some technical differences with the work of \cite{hanhart}, where the final $\pi\pi$ state interaction is unitarized using the Omn\`es representation, with dispersion relations. This approach requires a subtraction constant which is fitted to the ratio $\Gamma(D_{s1}(2460) \to D_s^+ \pi^+ \pi^-)/\Gamma(D_{s1}(2460) \to D_s^{*+} \pi^0)$. In contrast, we use directly the chiral unitary approach, where the cutoff regulating the loops is fitted to $\pi\pi$ scattering data.
Note also that the unitary $K\bar{K} \to \pi^+\pi^-$, $\eta\eta \to \pi^+\pi^-$ amplitudes only appear in the diagrams in Fig.~\ref{fig:diagtriang}(a), (b) and (c), while in diagrams (d) and (e) the upper pion in the loop appears in p-wave, while the lower one appears in s-wave, and there is no rescattering in p-wave in this limited phase space (the $\rho$ meson is very far away).

\begin{figure}[htbp]
\begin{center}
\includegraphics[width=0.45\textwidth]{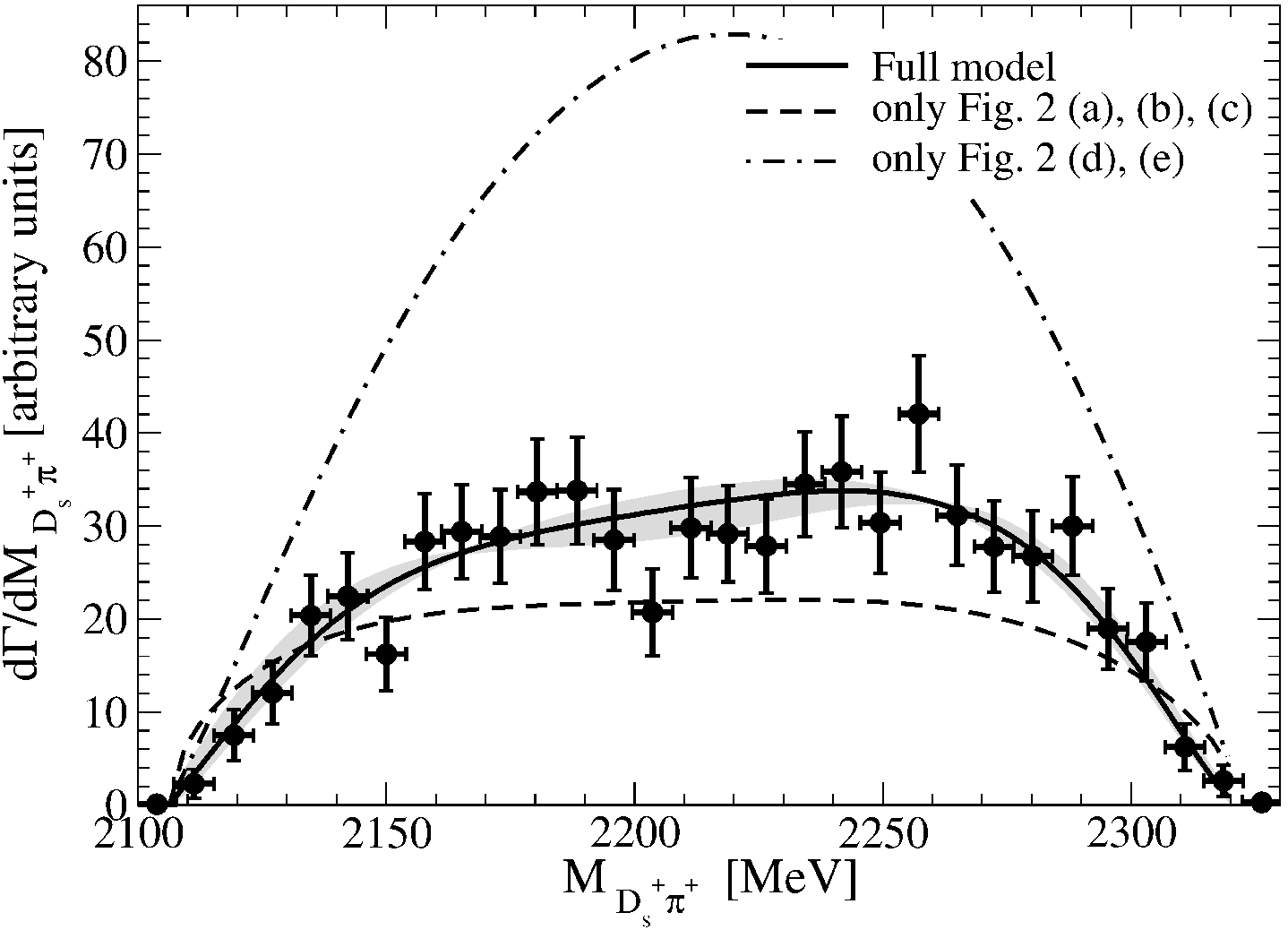}
\caption{\small{Comparison of the different contributions to the experimental $D_s^+\pi^+$ distribution \cite{expe}.}}
\label{fig:Minv_Dspi}
\end{center}
\end{figure}

In Fig.~\ref{fig:Minv_Dspi}, we show our results for the  $D_s^+\pi^+$ distribution compared to the experimental data from \cite{expe}, using the same normalization as in Fig.~\ref{fig:Minv_pipi}\footnote{In the experimental data from Ref.~\cite{expe}, the area of the $D_s^+\pi^+$ distribution is about  8\% smaller than that of the $\pi^+\pi^-$ distribution. Therefore we have increased the experimental data $D_s^+\pi^+$ by that amount in the plots.}. Once again, we see that the interference between diagrams \ref{fig:diagtriang} (a), (b), (c)  with (d),(e), is important in shaping the final distribution, but less critical than for the $\pi^+\pi^-$ distribution, resulting in overall good agreement with the experimental data.

At this point, we should mention that the contribution from diagram \ref{fig:diagtriang}~(c) is indeed small, as assumed in \cite{hanhart}. However, this is not a trivial result that one could have anticipated. Indeed, in our model, this suppression arises from a combination of factors, all contributing in the same direction. First, the meson-meson scattering amplitude leading to the final $\pi^+\pi^-$ state (represented by the thick dot in Fig.~\ref{fig:diagtriang}) is a factor about 4 to 8 times smaller (depending on the $\pi\pi$ energy)  for diagram (c) compared to (a) or (b). This is mostly due to the smaller tree level potential $V_{\eta\eta,\pi^+\pi^-}$ relative to $V_{K^+K^-,\pi^+\pi^-}$, and from the larger contribution in the latter case of the nearby $f_0(500)$ upon unitarization (manifest in its stronger coupling to $K^+K^-$ than to $\eta\eta$). 
Second, the $D_s^{*+} \to D_s^+ \eta$ vertex is suppressed by a factor of $1/\sqrt{3}$ compared to the $D_s^{+} \to D_s^+ \bar{K}$ vertex, as shown in Table~\ref{tab:beta}, and additionally the first vertex $D_{s1} \to D_s^* \eta$ is also a bit smaller than $D_{s1} \to D^* \bar{K}$. Furthermore, the triangle loop function $Y(D_s^+,\eta,\eta)$ is about a factor 2.5 times smaller than $Y(D^*,\bar{K},K)$. Finally, there is an additional factor of two, as there are two diagrams of the $\bar{K}$ exchange type, (a) and (b). Altogether all these contributions combine to give an amplitude for diagram (c) which is about 50 to 90 times smaller than the contribution from (a) and (b).

\begin{figure}[h]
\begin{center}
\includegraphics[width=0.45\textwidth]{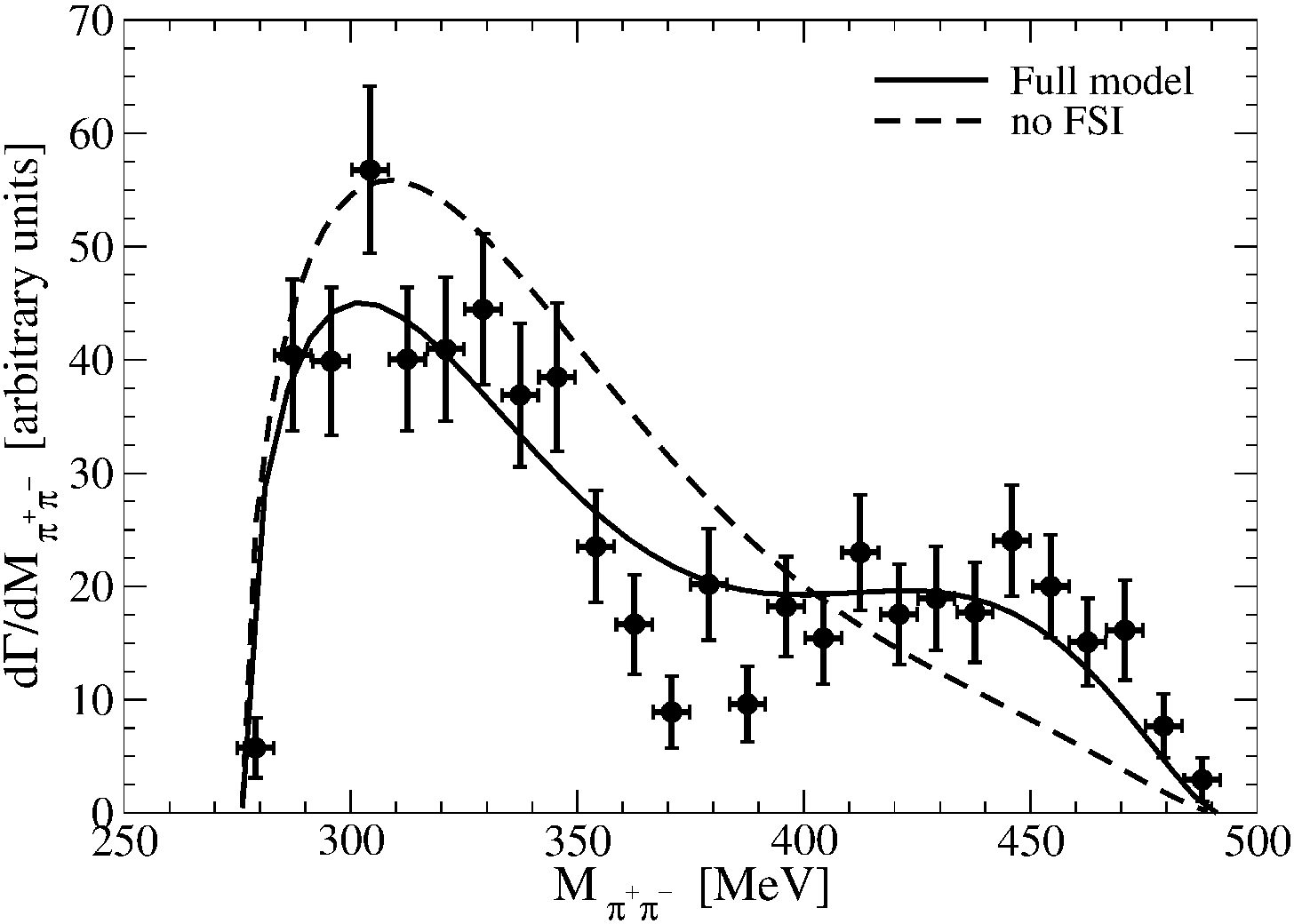}
\caption{Effect of removing the final state interaction in the meson-meson unitarized scattering amplitudes.}
\label{fig:fignoFSI}
\end{center}
\end{figure}

Returning to the discussion on the contribution of the scalar resonances $f_0(500)$ and $f_0(980)$, we show in Fig.~\ref{fig:fignoFSI} the $\pi^+\pi^-$ mass distribution but removing the final state interaction in coupled channels that lead to the final $\pi^+\pi^-$ state. That is, using just the tree level potential $V_{K^+K^-,\pi^+\pi^-}$ instead of the unitarized scattering amplitude $t_{K^+K^-,\pi^+\pi^-}$ in Eq.~\eqref{eqtabc}, and similarly for the other channels. In this way, the scalar resonances never get to be generated, particularly the $f_0(500)$ which is the most relevant in the present case. We see in the figure that the effect of this meson-meson final state interaction, namely the $f_0(500)$, is to produce the double bump structure via interference with the diagrams \ref{fig:diagtriang}(d) and 
\ref{fig:diagtriang}(e).

Finally, our approach allows us to calculate the total decay width for $D_{s1}^+ \to D_s^+ \pi^+ \pi^-$,  not provided in \cite{expe}, for which we find
\begin{equation}
    \Gamma(D_{s1}^+ \to D_s^+ \pi^+ \pi^-) = 4.3\pm 1.2\text{ keV}.
\end{equation}
where the error comes again from the most important source of uncertainty, namely the cutoff $\Lambda'$.

While the shape and the effects of the final state interaction are very similar to those of Ref.~\cite{hanhart}, the integrated width is somewhat different. In Ref.~\cite{hanhart} the $D_{s1} \to D_s \pi^+ \pi^- $ width obtained is $16^{+7}_{-5}$~keV, still at least a factor two larger than ours considering errors.
 One reason for our smaller width is the implementation of the cutoff in the $K \bar K\to \pi \pi$ and $K D \to \pi D_s$ amplitudes, which is inherent to the chiral unitary approach.
Removing this cutoff would lead to an increase in the total width by about 35\%. The differences in the total width between Ref.~\cite{hanhart} and the present work should be looked at in a broader context by mentioning results from other works. Indeed, in Ref.~\cite{bardeen}, using $SU(3)_L\times SU(3)_R$ chiral symmetry, a width $\Gamma=1.9$~keV is obtained. In Ref.~\cite{45}, using arguments of heavy quark effective theory and loop mechanisms, the width obtained is $\Gamma=0.25$~keV. It would be interesting to have the total width measured to put further constraints on the theoretical approaches.

As a conclusion of the results section, we must say that assuming the molecular picture for $D_{s1}(2460)$ we achieve a good reproduction of the mass distributions observed in the LHCb experiment without any free parameters. This remarkable fact is certainly a strong element in support of the molecular picture for this resonance.

\section{Conclusions}

We perform a theoretical study of the $D_{s1} \to D_s^+ \pi^+ \pi^-$ reaction from the perspective that the $D_{s1}$ is a molecular state mostly made from the $D^* K$ and $D_s^* \eta$ components, as found in several independent works. 
In this picture, the $D^*$ or $D_s^*$ states decay to two pseudoscalars, thus leading to the formation of three pseudoscalar mesons. However, none of the possible combinations of these three mesons correspond directly to the final $D_s^+ \pi^+ \pi^-$ state, thus requiring additional interactions to reach that final state. In the next step, two of these mesons merge, allowing them to interact, and produce a different pair of mesons leading to the final $D_s^+ \pi^+ \pi^-$ state through mechanisms involving a triangle loop. 
The meson-meson interaction is obtained implementing the techniques of the chiral unitary approach or the local hidden gauge approach.
We compare the resulting $\pi^+ \pi^-$ and  $D_s^+ \pi^+$ invariant mass distributions to recent experimental data from the LHCb collaboration.
The agreement is remarkable, especially considering that our model contains no free parameters, once the interaction is tuned to meson-meson scattering data, and hence no fitting is performed to the experimental LHCb distributions.
This provides strong support to the molecular picture of the $D_{s1}$ resonance.

\section{Acknowledgments}

This work is supported by the Spanish Ministerio
de Economia y Competitividad (MINECO) and European FEDER funds under Contracts No. FIS2017-84038-C2-1-P B,
PID2020- 112777GB-I00, and by Generalitat Valenciana under contract PROMETEO/2020/023. This project has received
funding from the European Union Horizon 2020 research and innovation programme under the program H2020- INFRAIA2018-1, grant agreement No. 824093 of the STRONG-2020 project. This work is supported by the Spanish Ministerio de
Ciencia e Innovaci\'on (MICINN) under contracts PID2020-112777GB-I00, PID2023-147458NB-C21 and CEX2023-001292-S;
by Generalitat Valenciana under contracts PROMETEO/2020/023 and CIPROM/2023/59.

\end{document}